\documentclass[journal]{IEEEtran}
\UseRawInputEncoding
\usepackage{multirow}
\usepackage{hyperref}
\usepackage{amsfonts,amssymb}
\usepackage{mathrsfs}
\usepackage{makecell} 
\usepackage{color}
\usepackage{bm} 

\pdfoutput=1
\ifCLASSINFOpdf
   \usepackage[pdftex]{graphicx}

\else

   \usepackage[dvips]{graphicx}

\fi

\usepackage{amsmath}
\usepackage{graphicx}

\usepackage{array}

\ifCLASSOPTIONcompsoc
  \usepackage[caption=false,font=normalsize,labelfont=sf,textfont=sf]{subfig}
\else
  \usepackage[caption=false,font=footnotesize]{subfig}
\fi

\begin{document}

\title{Deep Learning Based Joint Space-Time-Frequency Domain Channel Prediction for Cell-Free Massive MIMO Systems}

\author{Yongning Qi, Tao Zhou,~\IEEEmembership{Senior Member,~IEEE,}
Zuowei Xiang,
Liu Liu,~\IEEEmembership{Member,~IEEE,}
and Bo Ai,~\IEEEmembership{Fellow,~IEEE}

\thanks{This work has been submitted to the IEEE for possible publication.}
\thanks{Y. Qi, T. Zhou, Z. Xiang, and L. Liu are with the School of Electronic and Information Engineering, Beijing Jiaotong University, Beijing 100044, China (e-mail: yongningqi@bjtu.edu.cn; taozhou@bjtu.edu.cn; 21125104@bjtu.edu.cn; liuliu@bjtu.edu.cn).}
\thanks{B. Ai is with the State Key Laboratory of Advanced Rail Autonomous Operation, Beijing 100044, China, and also with the School of Electronic and Information Engineering, Beijing Jiaotong University, Beijing 100044, China (e-mail: boai@bjtu.edu.cn).}
}


\maketitle

\begin{abstract}
The cell-free massive multi-input multi-output (CF-mMIMO) is a promising technology for the six generation (6G) communication systems. Channel prediction will play an important role in obtaining the accurate CSI to improve the performance of CF-mMIMO systems. This paper studies a deep learning (DL) based joint space-time-frequency domain channel prediction for CF-mMIMO. Firstly, the prediction problems are formulated, which can output the multi-step prediction results in parallel without error propagation. Then, a novel channel prediction model is proposed, which adds frequency convolution (FreqConv) and space convolution (SpaceConv) layers to Transformer-encoder. It is able to utilize the space-time-frequency correlations and extract the space correlation in the irregular AP deployment. Next, simulated datasets with different sizes of service areas, UE velocities and scenarios are generated, and correlation analysis and cross-validation are used to determine the optimal hyper-parameters. According to the optimized hyper-parameters, the prediction accuracy and computational complexity are evaluated based on simulated datasets. It is indicated that the prediction accuracy of the proposed model is higher than traditional model, and its computational complexity is lower than traditional Transformer model. After that, the impacts of space-time-frequency correlations on prediction accuracy are studied. Finally, realistic datasets in a high-speed train (HST) long-term evolution (LTE) network are collected to verify the prediction accuracy. The verification results demonstrate that it also achieves higher prediction accuracy compared with traditional models in the HST LTE network.
\end{abstract}


\IEEEpeerreviewmaketitle

\section{Introduction}
\IEEEPARstart{C}ELL-FREE massive multi-input multi-output (CF-mMIMO) is one of potential technologies for the sixth generation (6G) communication systems \cite{10522673}. Compared with traditional centralized mMIMO, it has more flexibility in antenna deployment and enables to provide greater service for user equipments (UEs) \cite{6965818,9944695,9103348}. The CF-mMIMO system consists of several central processing units (CPUs) and a large number of access points (APs) with one or more antennas scattered in a coverage area. When UEs move at a high speed, serving clusters defined as subsets of available APs need to be reassigned frequently, which will affect the system performance \cite{9650567}. The performance of CF-mMIMO systems mainly hinges on the accuracy of instantaneous channel state information (CSI). However, acquiring accurate CSI poses a significant challenge, as conventional channel estimation methods suffer from excessive pilot overhead, rendering them fail to keep pace with the rapid time-varying of channels. In contrast to the channel estimation, channel prediction leverages historical CSI to forecast the future CSI for a specific time span without pilot overhead. This approach mitigates the detrimental effects of processing delays and reduces the impact of outdated CSI, thereby enhancing the performance of CF-mMIMO systems. \cite{4389753}.

Previously, traditional statistical models, such as an autoregressive (AR) model and a parametric model, were commonly employed for channel prediction \cite{1512123,6945858}. Nevertheless, for complex rapid time-varying channels, the statistical models often are difficult to reach a high prediction accuracy \cite{8813020}. Nowadays, artificial intelligence (AI) technology, with exceptional capability for feature extraction, has experienced remarkable advancements over the past decade \cite{9844014,9210812}. With the rapid popularization of AI, deep learning (DL), recognized as a pivotal subset of AI, is also increasingly applied for the channel prediction to deliver the superior prediction accuracy \cite{8761308,aibo6grailway}. Besides, it can also accelerate the evolution of the 6G communication networks into a kind of AI-native networks, where AI technologies are utilized at the very beginning of network design, thereby embedding inherent intelligence in the network architecture \cite{9952578,10062570}. 

For single-input single-output (SISO) systems, the channel prediction usually considers the time correlation between CSI at different moments. In \cite{6755477}, a complex-valued neural network (CVNN) combined with a chirp Z-transform was resorted to forecast the CSI by utilizing the time correlation. It is able to deal with complex values directly, which has high flexibility and decreases the generalization error in the CSI prediction. Similarly, a long short-term memory (LSTM) network with selective memory function was used to predict the CSI under the Rayleigh channel, which can filter the effective time correlation through forget gates, input gates and output gates, and store the relevant information in the memory cell \cite{9002073}. A sequence-to-sequence (Seq2Seq) model incorporating LSTM cells was developed for CSI prediction \cite{8904286}, and its performance was validated through datasets collected in outdoor scenarios. While the Seq2Seq model can break through the limitation of capturing global time correlation, it remains prone to gradient vanishing and error accumulation when processing long time sequences. In order to address these issues, Transformer proposed by Google utilized the global time correlation to predict the CSI through multi-head attention mechanism \cite{NIPS2017_3f5ee243,10480335}. In this novel attention mechanism, several attention heads can analyze the input series from different perspectives, thereby enhancing the capability for capturing the global time correlation. Nevertheless, the above-mentioned researches predominantly focus on leveraging the time correlation for channel prediction in narrowband SISO systems. In contrast, for wideband SISO systems, a CSI prediction model incorporating both time and frequency correlations was proposed in \cite{cnntransformer}. This model integrates a convolutional neural network (CNN) with dual Transformer encoders to jointly extract time and frequency features, enabling time-frequency joint wideband CSI prediction.

Beyond SISO systems, the channel prediction has also been extensively applied in MIMO systems. A deep recurrent neural network (RNN) incorporating LSTM or gated recurrent unit (GRU), which captures the time correlation through memory cell, was used to predict the CSI for distributed flat-fading MIMO channels \cite{9044427}. In \cite{10185566}, an adaptive bidirectional GRU (ABiGRU), as a novel model, was employed to predict the CSI by utilizing the time correlation for underwater acoustic MIMO systems. This model effectively relieves the error propagation and achieves more accurate time correlation extraction. The research in \cite{9832933} used a Transformer architecture, incorporating a self-attention mechanism, for CSI prediction in centralized mMIMO systems. In this model, the self-attention mechanism can assign different weights of CSI at different moments to improve the ability for the time correlation extraction. In \cite{10529153}, a Transformer variant called Gruformer is proposed to forecast the CSI at the transmitter for MIMO systems of vehicle-to-everything (V2X). This hybrid architecture merges GRU modules with the Transformer framework to effectively capture local time correlation between adjacent CSI sampling points. However, these above-mentioned researches only consider the time correlation between the CSI at different moments, but they do not consider the space correlation between antennas. 

In addition to utilizing the time correlation, the space correlation is also leveraged to achieve CSI prediction. As an example, a multi-layer perceptron (MLP) architecture, comprising stacked dense layers, was designed to utilize the space correlation for accurate CSI prediction in massive MIMO systems, while maintaining minimal computational overhead during online training \cite{10647106}. In order to further improve the prediction accuracy, there are several studies for CSI prediction, which consider both space and time correlations simultaneously. A GRU was used to predict the CSI for the MIMO systems by extracting the space-time correlations \cite{10946619}. A novel model based on a convolutional LSTM (CLSTM) network for CSI prediction in the massive MIMO systems, which considers the space-time correlations \cite{9921297}. A spatial attention layer, which consists of two pooling layers and a convolutional layer, is able to embed the space feature in the MIMO CSI datasets in this approach. Similarly, a CNN and a CLSTM network were combined to propose a novel hybrid DL model called Conv-CLSTM, which leverages space-time correlations to predict the narrowband CSI for the centralized mMIMO systems \cite{9676455}. The two-dimensional convolution kernels, integrated into the Conv-CLSTM, enable to extract the space correlation by sliding in the space dimension. However, the CNN is typically used to extract the space correlation for datasets with regular space structure, but it has a limitation in extracting the space correlation for datasets with irregular space structure. Although existing researches have laid an important foundation for CSI prediction, there is still a lack of reseach on wideband CSI prediction for CF-mMIMO systems utilizing space-time-frequency correlations. To fill this research gap, two main challenge should be concerned, involving how to extract the space correlation for the CF-mMIMO systems with irregular AP deployment and how to implement the joint space-time-frequency domain channel prediction.

This paper aims to investigate the joint space-time-frequency channel prediction for the CF-mMIMO systems based on DL models. The key contributions and innovations in this paper are as follows:

1) The channel prediction problems for wideband CF-mMIMO systems are formulated, which consider the space-time-frequency correlations simultaneously. Crucially, different from traditional recursive prediction problems which perform single-step ahead prediction sequentially, the prediction problems can predict the multi-step CSI in the future at the same time without error propagation. 

2) A novel channel prediction model is proposed, which enhances the Transformer-encoder architecture through adding the frequency convolution (FreqConv) and space convolution (SpaceConv) layers. This design can capture the multi-dimensional correlations to realize the joint space-time-frequency domain channel prediction, and address the challenge for space correlation extraction in the irregular AP deployment for CF-mMIMO systems. Moreover, its space and time complexity are reduced compared with the vanilla Transformer.

3) The simulated datasets are generated to evaluate the prediction accuracy and computational complexity. Based on these datasets, ablation experiment is implemented to prove the effectiveness of the FreqConv and SpaceConv layers. Then, we compare the prediction accuracy and computational complexity of the proposed model and traditional DL models. These indicate that the proposed model has the best prediction accuracy compared with traditional DL models, and the computational complexity decreases compared with the vanilla Transformer. Finally, the influences of space-time-frequency correlations on prediction accuracy are studied.

4) The realistic channel data is measured in a high-speed train (HST) long-term evolution (LTE) network. Utilizing a time-delay-window-based partitioning method, the CSI datasets are derived from the realistic channel data. After that, the prediction accuracy of the proposed model is verified based on the realistic datasets. It is indicated that the proposed model still shows the best prediction accuracy in the HST LTE network compared with traditional DL models.

The remainder of this paper is outlined as follows. Section II introduces the method of prediction problem formulation. In Section III, a joint space-time-frequency domain channel prediction model for wideband CF-mMIMO systems based on DL is proposed. Section IV describes the performance evaluation of the proposed model based on simulated datasets. Similarly, the performance validation is performed using datasets measured in the HST LTE network in Section V. Finally, conclusions are drawn in Section VI.

\section{Prediction Problem Formulation}
We consider a single-user CF-mMIMO system with $L$ subcarriers. It comprises a CPU and $M$ APs, where the APs are randomly distributed in a square area and connected to the CPU via fronthaul links. All APs and the UE take place over the same frequency resource, and are equipped with single antenna, respectively. For this system, the received signal of the UE from $m$-th AP at time $t$ can be expressed as
\begin{equation}
\bm{\mathrm{y}}_{m,~t}=\sqrt{P}\sum_{m=1}^M\bm{\mathrm{h}}_{m,~t}s_t+\bm{\mathrm{n}}_{m,~t} \text{,}
\end{equation}
where $P$ is the transmitted power of the UE, $s_t$ represents the transmitted signal, $\bm{\mathrm{n}}_{m,~t} \in \mathbb{C}^{1\times L}$ indicates the additive white Gaussian noise (AWGN) vector, which is assumed to be a random vector following Gaussian distribution $\mathcal{CN}(0,\sigma_n^2)$ with $\sigma_n^2$ as the power density of noise, $\bm{\mathrm{h}}_{m,~t}=\left(h_{m,~t}^1,\cdots,h_{m,~t}^l,\cdots,h_{m,~t}^L\right)^T\in\mathbb{C}^{1\times L}$ stands for the CSI between the $m$-th AP and the UE at time $t$, and $h_{m,~t}^l\in\mathbb{C}$ is the CSI of the $l$-th subcarrier between the $m$-th AP and the UE at time $t$. 

Hence, the CSI $\bm{\mathrm{H}}_t\in \mathbb{C}^{L \times M}$ between $M$ APs and the UE at time $t$ can be written as
\begin{equation}\boldsymbol{\mathrm{H}}_{t}=\begin{bmatrix}h_{1,~t}^{1}&h_{2,~t}^{1}&\cdots&h_{M,~t}^{1}\\h_{1,~t}^{2}&h_{2,~t}^{2}&\cdots&h_{M,~t}^{2}\\\vdots&\vdots&\ddots&\vdots\\h_{1,~t}^{L}&h_{2,~t}^{L}&\cdots&h_{M,~t}^{L}\end{bmatrix}\text{.} \end{equation}
Since the $\bm{\mathrm{H}}_t$ is a complex-valued matrix, it is separated from real part and imaginary part for adapting to the DL models, which are represented as $\bm{\mathrm{H}}_t^\mathrm{(r)}=\Re\left(\bm{\mathrm{H}}_t\right)\in\mathbb{R}^{L\times M}$ and $\bm{\mathrm{H}}_t^\mathrm{(i)}=\Im\left(\bm{\mathrm{H}}_t\right)\in\mathbb{R}^{L\times M}$, respectively. Furthermore, the CSI in the past $T$ moments may contain the space, time and frequency correlations, simultaneously. Taking the real part of CSI as an example, to begin with, the space and frequency correlations are extracted, as follows
\begin{equation}
\tilde{\boldsymbol{\mathrm{H}}}_t^\mathrm{(r)}=\left[\mathcal{S}\left(\boldsymbol{\mathrm{H}}_t^\mathrm{(r)}\right),\mathcal{F}\left(\boldsymbol{\mathrm{H}}_t^\mathrm{(r)}\right)\right] \text{,}
\end{equation}
where $\mathcal{S}(\cdot)$ and $\mathcal{F}(\cdot)$ represent the extractors for space and frequency correlations, respectively. 

\begin{figure*}[t]
  \includegraphics[trim=15pt 15pt 30pt 15pt,clip,scale=0.75]{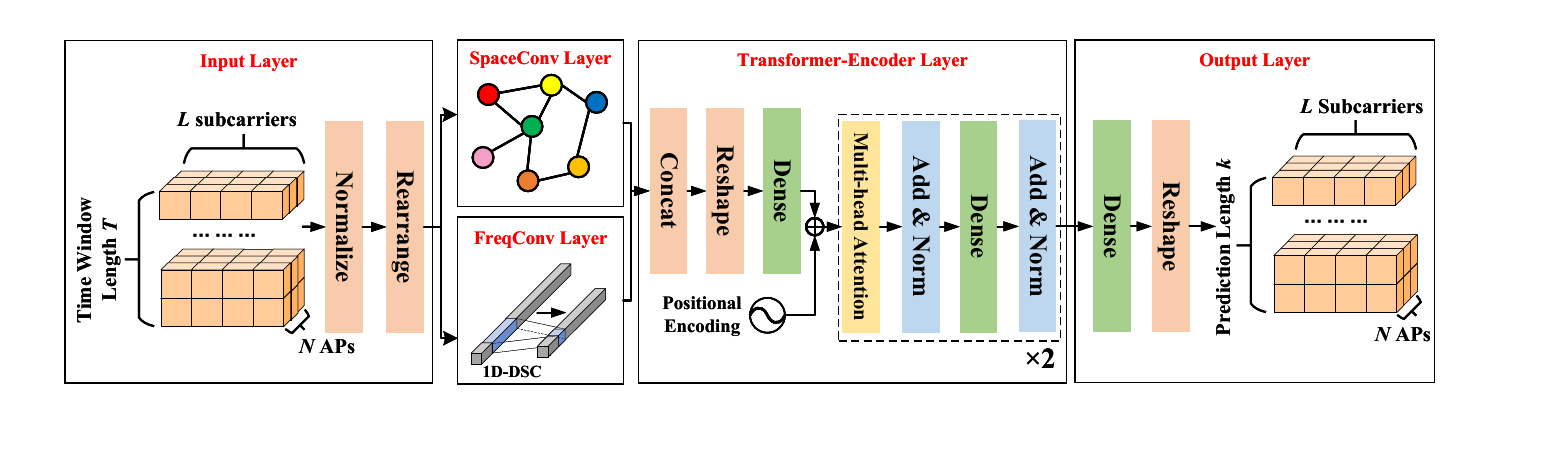}
  \caption{The overall structure of the proposed model. \label{fig:1}}
\end{figure*}
After that, we utilize the historical CSI processed by $\mathcal{S}(\cdot)$ and $\mathcal{F}(\cdot)$ in the past $T$ moments $\{\tilde{\boldsymbol{\mathrm{H}}}_{t-T+1}^\mathrm{(r)}$,$\cdots$,$\tilde{\boldsymbol{\mathrm{H}}}_{t}^\mathrm{(r)}\}$ to predict the future CSI in the next $K$ moments $\{\hat{\boldsymbol{\mathrm{H}}}_{t+1}^\mathrm{(r)},\cdots,\hat{\boldsymbol{\mathrm{H}}}_{t+K}^\mathrm{(r)}\}$. The normalized mean square error (NMSE) serves as a key metric for evaluating prediction accuracy by quantifying the relative magnitude of prediction errors. Compared with MSE affected by the amplitude of CSI, the NMSE could eliminate this effect by normalizing. The smaller the NMSE is, the higher the prediction accuracy of model will be. Hence, the channel prediction problems are formulated as
\begin{subequations}
  \begin{align}
    &\min_\Theta\mathcal{L}(\Theta)=\mathbb{E}\left\{\frac{\sum_{k=1}^K\left\|\hat{\boldsymbol{\mathrm{H}}}_{t+k}^\mathrm{(r)}-\boldsymbol{\mathrm{H}}_{t+k}^\mathrm{(r)}\right\|^2}{\sum_{k=1}^K\left\|\boldsymbol{\mathrm{H}}_{t+k}^\mathrm{(r)}\right\|^2}\right\}\text{,}\\
    &s.t.\quad\left(\hat{\boldsymbol{\mathrm{H}}}_{t+1}^\mathrm{(r)},\cdots,\hat{\boldsymbol{\mathrm{H}}}_{t+K}^\mathrm{(r)}\right)=\mathcal{T}_{\Theta}\left(\tilde{\boldsymbol{\mathrm{H}}}_{t-T+1}^\mathrm{(r)},\cdots,\tilde{\boldsymbol{\mathrm{H}}}_{t}^\mathrm{(r)}\right)\text{,}
  \end{align}
\end{subequations}
where $\hat{\boldsymbol{\mathrm{H}}}_{t+K}^\mathrm{(r)}\in\mathbb{R}^{L\times M}$ represents the real part of prediction results at time $t+K$, $\mathcal{T}_{\Theta}(\cdot)$ denotes the extractor for time correlation, and $\Theta$ is the option of hyper-parameters for the DL model. 

\section{A Joint Space-Time-Frequency Domain Channel Prediction Model}
\subsection{Overall Structure}

Prior researches demonstrate that traditional DL-based channel prediction models typically consider either one-dimensional correlation or two-dimensional correlations, such as time correlation, frequency correlation and space-time correlations, in centralized mMIMO systems with regular space structure. In this section, we propose a joint space-time-frequency domain channel prediction model for wideband CF-mMIMO systems. This proposed model advances beyond conventional approaches by exploiting space, time and frequency correlations, simultaneously. Notably, it possesses the capability for extracting the space correlation hidden in the CSI datasets of CF-mMIMO systems characterized by irregular space structure.

As illustrated in Fig.~\ref{fig:1}, the overall architecture of the proposed model includes an input layer, a SpaceConv layer, a FreqConv layer, a Transformer-encoder layer, and an output layer. The input layer normalizes the input datasets to conform to a standard normal distribution, and executes the dimension arrangement to transform the raw datasets into the appropriate tensor format required for subsequent neural network layer processing. The SpaceConv layer leverages the spectral convolution to extract the space correlation in datasets with irregular space structure. It is able to effectively overcome the limitation of traditional convolution, which can only extract local correlation in regularly structured datasets. The FreqConv layer employs a one-dimensional depthwise separable convolution (1D-DSC) to capture the local frequency correlation. This innovation design achieves to improve the computational efficiency, thereby enhancing real-time processing capability. The Transformer-encoder layer mainly comprises two multi-head attention sub-layers, two dense sub-layers, and four addition and layer normalization (Add \& Norm) sub-layers. In this layer, the outputs from the FreqConv and SpaceConv layers are concatenated, and fed into the multi-head attention sub-layer after reshaping. The multi-head attention mechanism can capture the time correlation between two moments in parallel through multi-head collaboration. Besides, the Add \& Norm sub-layer and the dense sub-layer can enhance the training stability by mitigating the issues of gradient vanishing or exploding. The output layer employs a dense sub-layer, which can adjust the dimension to output the prediction results at several moments through reshape operation.

\subsection{SpaceConv Layer}

\begin{figure}[t]
  \includegraphics[trim=15pt 15pt 15pt 15pt,clip,scale=1]{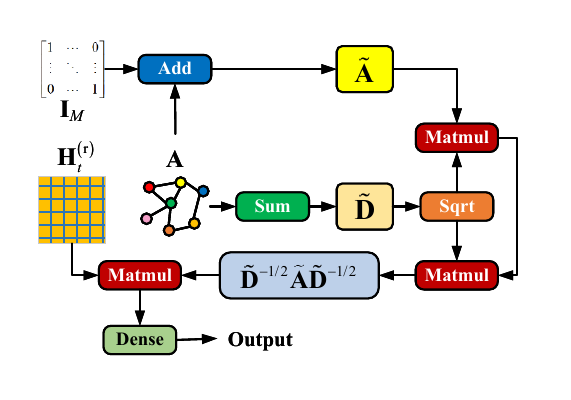}
  \caption{The structure of the SpaceConv layer. \label{fig:SpaceConv}}
  \vspace{-0.45cm}
\end{figure}

The SpaceConv layer employs the spectral convolution to extract the space correlation in the CSI of CF-mMIMO systems regarded as a weighted graph, where the APs and the correlation values correspond to nodes and weights of edges, respectively. It can update the features of nodes through information passing and aggregation between nodes, thereby enabling it to extract the space correlation in the datasets with irregular space structure directly \cite{graphconvolutionnetwork}. 

The spectral convolution is defined as the multiplication of the CSI at time $t$ with a parameterized filter $g_f(\cdot)$ in the Fourier domain, expressed as
\begin{equation}
  \mathcal{S}\left(\boldsymbol{\mathrm{H}}_t^\mathrm{(r)}\right)=\boldsymbol{\mathrm{U}}g_f(\boldsymbol{\Lambda})\left(\boldsymbol{\mathrm{H}}_t^\mathrm{(r)}\boldsymbol{\mathrm{U}}\right)^\top\text{,}
\end{equation}
where $\bm{\mathrm{U}}\in\mathbb{R}^{M\times M}$ represents the eigenmatrix of the normalized graph Laplacian, $\bm{\Lambda}\in\mathbb{R}^{M\times M}$ denotes the diagonal matrix of its eigenvalues, and $\bm{\mathrm{L}}\in\mathbb{R}^{M\times M}$ is the normalized graph Laplacian, calculated as
\begin{equation}
  \bm{\mathrm{L}}=\bm{\mathrm{I}}_M-\bm{\mathrm{D}}^{-1/2}\bm{\mathrm{A}}\bm{\mathrm{D}}^{-1/2} \text{,}
\end{equation}
where $\bm{\mathrm{I}}_M\in\mathbb{R}^{M\times M}$ indicates the $M$-order identity matrix, $\bm{\mathrm{D}}\in\mathbb{R}^{M\times M}$ represents the degree matrix, and $\bm{\mathrm{A}}\in\mathbb{R}^{M\times M}$ is the adjacency matrix.

To reduce the computational complexity of this layer, $g_f(\bm{\Lambda})$ can be effectively approximated by a truncation of Chebyshev polynomials $\mathcal{T}_k(\cdot)$ up to the $k$-th order, expressed as
\begin{subequations}
  \begin{align}
    \mathcal{S}\left(\boldsymbol{\mathrm{H}}_t^\mathrm{(r)}\right)&\approx\sum_{k=0}^K\bm{\mathrm{W}}_\mathrm{s}^k\mathcal{T}_k(\tilde{\boldsymbol{\mathrm{L}}})\left(\boldsymbol{\mathrm{H}}_t^\mathrm{(r)}\right)^\top \text{,}\\
    s.t.\quad\tilde{\bm{\mathrm{L}}}&=\bm{\mathrm{L}}-\bm{\mathrm{I}}_M \text{,}
  \end{align}
\end{subequations}
where $\bm{\mathrm{W}}_\mathrm{s}^k\in \mathbb{R}^{M \times M}$ represents the weight matrix of the $k$-th order term in the Chebyshev polynomial, and $\tilde{\bm{\mathrm{L}}}\in\mathbb{R}^{M\times M}$ is the normalized graph Laplacian $\bm{\mathrm{L}}$ after rescaling. Usually, the order of Chebyshev polynomials is set to one, as follows
\begin{subequations}
  \begin{align}
     \mathcal{S}\left(\boldsymbol{\mathrm{H}}_t^\mathrm{(r)}\right)&=\bm{\mathrm{W}}_\mathrm{s}^0\mathcal{T}_0(\hat{\boldsymbol{\mathrm{L}}})\boldsymbol{\mathrm{H}}_t^\mathrm{(r)}+\bm{\mathrm{W}}_\mathrm{s}^1\mathcal{T}_1(\hat{\boldsymbol{\mathrm{L}}})\boldsymbol{\mathrm{H}}_t^\mathrm{(r)} \text{,}\\
     s.t.\quad \mathcal{T}_0(\hat{\bm{\mathrm{L}}})&=\bm{\mathrm{I}}_M \text{,}\quad \mathcal{T}_1\left(\hat{\bm{\mathrm{L}}}\right)=\hat{\bm{\mathrm{L}}} \text{,}
  \end{align}
\end{subequations}
where $\mathcal{T}_k(\cdot)$ stands for the $k$-th Chebyshev polynomials. Then, we set $\bm{\mathrm{W}}_s=\bm{\mathrm{W}}_\mathrm{s}^0=-\bm{\mathrm{W}}_\mathrm{s}^1$. Hence, the $\mathcal{S}\left(\boldsymbol{\mathrm{H}}_{t}^\mathrm{(r)}\right)$ is calculated as
\begin{equation}
  \begin{aligned}
    \mathcal{S}\left(\boldsymbol{\mathrm{H}}_{t}^\mathrm{(r)}\right)&=\bm{\mathrm{W}}_\mathrm{s}^0\boldsymbol{\mathrm{I}}_{M}\boldsymbol{\mathrm{H}}_{t}^\mathrm{(r)}+\bm{\mathrm{W}}_\mathrm{s}^1\boldsymbol{\hat{\mathrm{L}}}\boldsymbol{\mathrm{H}}_{t}^\mathrm{(r)}\\&=\bm{\mathrm{W}}_\mathrm{s}\boldsymbol{\mathrm{I}}_M\boldsymbol{\mathrm{H}}_t^\mathrm{(r)}-\bm{\mathrm{W}}_\mathrm{s}\left(\boldsymbol{\mathrm{L}}-\boldsymbol{\mathrm{I}}_M\right)\boldsymbol{\mathrm{H}}_t^\mathrm{(r)}\\&=\bm{\mathrm{W}}_\mathrm{s}\left(\bm{\mathrm{D}}^{-1/2}\bm{\mathrm{A}}\bm{\mathrm{D}}^{-1/2}+\bm{\mathrm{I}}_M\right)\boldsymbol{\mathrm{H}}_t^\mathrm{(r)} \text{.}
  \end{aligned}
\end{equation}
In order to avoid the loss of current node information and balance the weights between adjacent nodes to prevent abnormal gradients, the element $d_{ij}\in\mathbb{R}$ in the degree matrix $\bm{\mathrm{D}}\in\mathbb{R}^{M\times M}$ and adjacency matrix $\bm{\mathrm{A}}$ of the graph are required to normalize. The normalized adjacency matrix $\tilde{\bm{\mathrm{A}}}\in\mathbb{R}^{M \times M}$ and the element $\tilde{d}_{ij}\in\mathbb{R}$ in the normalized degree matrix $\tilde{\bm{\mathrm{D}}}\in\mathbb{R}^{M \times M}$ are expressed as 
\begin{equation}
  \tilde{\bm{\mathrm{A}}}=\bm{\mathrm{A}}+\bm{\mathrm{I}}_M \text{,}
\end{equation}
\begin{equation}
  \tilde{d}_{ij}=\sum_ia_{ij} \text{,}
\end{equation}
where $a_{ij}\in\mathbb{R}$ denotes the element in the adjacency matrix. Based on the above analysis, the operation process of the SpaceConv layer is shown in Fig.~\ref{fig:SpaceConv}, and the formula can be written as
\begin{equation}
  \mathcal{S}\left(\boldsymbol{\mathrm{H}}_t^{(r)}\right)=\bm{\mathrm{W}}_s\left(\tilde{\boldsymbol{\mathrm{D}}}^{-1/2}\tilde{\boldsymbol{\mathrm{A}}}\tilde{\boldsymbol{\mathrm{D}}}^{-1/2}\right)\boldsymbol{\mathrm{H}}_t^{(r)} \text{.}
\end{equation}

\subsection{FreqConv Layer}
In contrast to the CSI in the space domain, the CSI in the frequency domain can be recognized as datasets with regular space structure. Since the local correlation is high in the frequency domain, the convolution kernel is selected to extract the frequency correlation. Moreover, it effectively decreases the computational complexity compared with the dense layer through sparse connectivity and weight sharing. To further reduce its computational complexity, we employs the DSC in the FreqConv layer, which consists of the DWC and the PWC. 

The DWC is used to apply a dedicated convolution kernel for each input feature map \cite{Efficientconvolutionalneuralnetworks}. As we regard the CSI at each moment as a feature map, it can be considered that different one-dimensional convolution kernels are utilized to extract the frequency correlation on CSI at different moments. The CSI processed by DWC at time $t$ can be expressed as
\begin{equation}
  \tilde{\boldsymbol{\mathrm{H}}}^{\prime\mathrm{(r)}}_t[i,j]=\sum_{k=0}^{D_k-1}\boldsymbol{\mathrm{W}}_t^\mathrm{DWC}[k,j]\cdot\boldsymbol{\mathrm{H}}^\mathrm{(r)}_t[i+k,j] \text{,}
\end{equation}
where $\boldsymbol{\mathrm{W}}^\mathrm{DWC}_t\in\mathbb{R}^{D_k \times M}$ indicates the weight matrix of convolution kernel at time $t$, and $D_k$ denotes the size of convolution kernel. The size of convolution kernel and the stride of convolution operation are the most important parameters in the DWC operation. Note that the stride is usually set to one. The size of convolution kernel affects the performance of correlation extraction. It can be determined through frequency correlation analysis, which will be introduced in the Section IV-B. The PWC is applied to compute a linear combination for the output of DWC via a convolution kernel with the size of $1 \times 1$, as follows
\begin{equation}
  \mathcal{F}\left(\boldsymbol{\mathrm{H}}_{t}^\mathrm{(r)}\right)=\boldsymbol{\mathrm{W}}^\mathrm{PWC}_t\left[\boldsymbol{\mathrm{H}}_{t-T+1}^{\prime\mathrm{(r)}},\cdots,\boldsymbol{\mathrm{H}}_{t}^{\prime\mathrm{(r)}}\right] \text{,}
\end{equation}
where $\bm{\mathrm{W}}^\mathrm{PWC}_t\in\mathbb{R}^{1 \times 1 \times T}$ denotes the weight matrix of PWC when extracting the frequency correlation in the CSI at time $t$.

\subsection{Transformer-Encoder Layer}
The Transformer model was utilized to extract the time correlation in the previous studies, which consists of an encoder and a decoder \cite{10480335,9832933,10529153}. Compared with RNN and LSTM, the Transformer integrates the multi-head attention mechanism to output the prediction results in parallel without error propagation and avoids the issues of gradient disappearance or explosion. However, as demonstrated in our prior work \cite{cnntransformer}, even without the decoder, the encoder alone is sufficient to capture the time correlation. Not only does it not lead to decrease the prediction accuracy, but it can also reduce the number of trainable parameters and avoid the risk of overfitting. Therefore, in this study, we adopt the encoder of Transformer as the core component for time correlation extraction.

When the outputs of the SpaceConv and FreqConv layers are input to this layer, they are concatenated to obtain the processed historical CSI $\tilde{\boldsymbol{\mathrm{H}}}_t^\mathrm{(r)}$, expressed as
\begin{equation}
  \tilde{\boldsymbol{\mathrm{H}}}_t^\mathrm{(r)}=\mathrm{Concat}\left(\mathcal{S}\left(\boldsymbol{\mathrm{H}}_t^\mathrm{(r)}\right),\mathcal{F}\left(\boldsymbol{\mathrm{H}}_t^\mathrm{(r)}\right)\right) \text{,}
\end{equation}
where $\mathrm{Concat}(\cdot)$ stands for the operation of matrix splicing. Then, the processed historical CSI undergoes the vectorization operation and linear transformation, as follows
\begin{equation}
  \tilde{\boldsymbol{\mathrm{h}}}_t^\mathrm{(r)}=\mathrm{vec}\left(\tilde{\boldsymbol{\mathrm{H}}}_t^\mathrm{(r)}\right) \bm{\mathrm{W}}_{d_{\text{model}}} \text{,}
\end{equation}
where $\mathrm{vec}(\cdot)$ represents the vectorization operation, $\bm{\mathrm{W}}_{d_{\text{model}}}\in\mathbb{R}^{2ML\times d_{\text{model}}}$ denotes the weight matrix in the linear transformation, and $d_{\text{model}}$ indicates the input dimension of multi-head attention sub-layer. 

In this model, the absence of recurrent or convolution operations inherently limits the capacity to capture the absolute position. To compensate this drawback, a positional encoding matrix $\bm{\mathrm{P}}\in\mathbb{R}^{T\times d_{\text{model}}}$ is introduced to inject the position information in the time sequence \cite{9710917}. Due to the advantages of periodicity, continuity and orthogonality, trigonometric functions can process the time series with random length, capture the differences and avoid the irrelevant information. Hence, the matrix $\bm{\mathrm{P}}$ is calculated as
\begin{equation}
  \boldsymbol{\mathrm{P}}[i,j]=\sin\left(\frac{i}{10000^{(j-j\mathrm{mod}2)/d_{\text{model}}}}-\frac{(j\mathrm{mod}2)\pi}{2}\right) \text{.}
\end{equation}
In order to embed the position information without disrupting the space-time-frequency correlations in the historical CSI in previous $T$ moments, denoted as $\tilde{\boldsymbol{\mathrm{H}}}^\mathrm{(r)}=\left[\tilde{\boldsymbol{\mathrm{h}}}_{t-T+1}^\mathrm{(r)},\cdots,\tilde{\boldsymbol{\mathrm{h}}}_{t}^\mathrm{(r)}\right]\in\mathbb{R}^{T\times d_{\text{model}}}$, the matrix $\bm{\mathrm{P}}$ and $\tilde{\bm{\mathrm{H}}}^\mathrm{(r)}$ should be summed, which can be represented as $\tilde{\boldsymbol{\mathrm{H}}}^\mathrm{(r)}+\alpha\boldsymbol{\mathrm{P}}$, where $\alpha$ is an adjustable coefficient. 

The core sub-layer of the Transformer-encoder layer is the multi-head attention sub-layer, which greatly is capable of capturing diverse features compared with the self-attention mechanism. In this sub-layer, the input $\tilde{\boldsymbol{\mathrm{H}}}^\mathrm{(r)}+\alpha\boldsymbol{\mathrm{P}}\in\mathbb{R}^{T\times d_{\text{model}}}$ is linearly transformed into $h$ different representations, which are called attention heads. Each attention head has three inputs, including query matrix $\bm{\mathrm{Q}}$, key matrix $\bm{\mathrm{K}}$ and value matrix $\bm{\mathrm{V}}$, and we compute the inputs of $i$-th attention head as
\begin{subequations}
  \begin{align}
    &\bm{\mathrm{Q}}_i=\left(\tilde{\boldsymbol{\mathrm{H}}}^\mathrm{(r)}+\alpha\boldsymbol{\mathrm{P}}\right)\bm{\mathrm{W}}_i^{\bm{\mathrm{Q}}} \text{,}\\
    &\bm{\mathrm{K}}_i=\left(\tilde{\boldsymbol{\mathrm{H}}}^\mathrm{(r)}+\alpha\boldsymbol{\mathrm{P}}\right)\bm{\mathrm{W}}_i^{\bm{\mathrm{K}}} \text{,}\\
    &\bm{\mathrm{V}}_i=\left(\tilde{\boldsymbol{\mathrm{H}}}^\mathrm{(r)}+\alpha\boldsymbol{\mathrm{P}}\right)\bm{\mathrm{W}}_i^{\bm{\mathrm{V}}} \text{,}
  \end{align}
\end{subequations}
where $\bm{\mathrm{W}}_i^{\bm{\mathrm{Q}}}\in\mathbb{R}^{d_{\text{model}}\times d_k}$, $\bm{\mathrm{W}}_i^{\bm{\mathrm{K}}}\in\mathbb{R}^{d_{\text{model}}\times d_k}$ and $\bm{\mathrm{W}}_i^{\bm{\mathrm{V}}}\in\mathbb{R}^{d_{\text{model}}\times d_v}$ are the weight matrices in the $i$-th attention head, and $d_k$ and $d_v$ denote the dimension of $\bm{\mathrm{K}}$ and $\bm{\mathrm{V}}$, respectively. Subsequently, the inputs of each attention head are preocessed by the scaled dot product attention operation, and the result of $i$-th attention head produces a normalized probability distribution via a softmax function $\mathrm{softmax}(\bm{x})=\exp(x_i)/\sum_i\exp(x_i)$, shown as
\begin{equation}
\boldsymbol{\mathrm{HEAD}}_i=\mathrm{softmax}\left(\frac{\bm{\mathrm{Q}}_i\bm{\mathrm{K}}_i^\top}{\sqrt{d_k}}\right)\bm{\mathrm{V}}_i\text{.}
\end{equation}
Finally, the results of $h$ attention heads are concatenated and multiplied with a weight matrix $\bm{\mathrm{W}}^{\bm{\mathrm{O}}}\in\mathbb{R}^{hd_v\times d_{\text{model}}}$ to obtain the output of multi-head attention sub-layer $\bm{\mathrm{O}}\in\mathbb{R}^{T\times d_{\text{model}}}$, as follows
\begin{equation}
  \bm{\mathrm{O}}=\mathrm{Concat}(\bm{\mathrm{HEAD}}_1,\cdots,\bm{\mathrm{HEAD}}_h)\bm{\mathrm{W}}^{\bm{\mathrm{O}}}\text{.}
\end{equation}

The Add \& Norm sub-layer is utilized to avoid the gradient explosion or disappearance and shorten the training time. It adds $\tilde{\boldsymbol{\mathrm{H}}}^\mathrm{(r)}+\alpha\boldsymbol{\mathrm{P}}$ to the output of multi-head attention sub-layer $\bm{\mathrm{O}}$ and normalizes the term $\tilde{\boldsymbol{\mathrm{H}}}^\mathrm{(r)}+\alpha\boldsymbol{\mathrm{P}}+\boldsymbol{\mathrm{O}}$, shown as
\begin{subequations}
  \begin{align}
    &\boldsymbol{\mathrm{Z}}=\mathrm{LN}\left(\tilde{\boldsymbol{\mathrm{H}}}^\mathrm{(r)}+\alpha\boldsymbol{\mathrm{P}}+\boldsymbol{\mathrm{O}}\right) \text{,}\\
    & s.t.\quad\mathrm{LN}(\bm{\mathrm{X}})=\frac{\bm{\mathrm{X}}[i,j]-u_{j}}{\sqrt{\delta_{j}^{2}+\varepsilon}}\text{,}
  \end{align}
\end{subequations}
where $\varepsilon$ is a small constant, which can avoid the ill-conditioned problem, and $u_j$ and $\delta_j^2$ denote the mean and variance of the $\bm{\mathrm{X}}[:,j]$, respectively. Finally, a dense layer is considered to further extract and synthesize the time correlation, denoted by
\begin{subequations}
  \begin{align}
    \bm{\mathrm{Y}}=\mathrm{LN}&(\bm{\mathrm{Z}}+\mathrm{Dense}(\bm{\mathrm{Z}}))\text{,}\\
    s.t.\quad\mathrm{Dense}(\bm{\mathrm{X}})&=\max\left(0,\bm{\mathrm{W}}_\text{d}^1\bm{\mathrm{X}}\right)\bm{\mathrm{W}}_\text{d}^2 \text{,}
  \end{align}
\end{subequations}
where $\bm{W}_d^1$ and $\bm{W}_d^2\in\mathbb{R}^{d_{\text{model}}\times d_{\text{model}}}$ are the weight matrices in the dense sub-layer.

\subsection{Training process}
The training process starts from an initial state where all weights are randomly selected and follow $\mathcal{CU}[-\sqrt{6}/\sqrt{n_k+n_{k+1}},\sqrt{6}/\sqrt{n_k+n_{k+1}}]$ \cite{understanddifficulty}, where $n_k$ represents the number of neurons in $k$-th layer, and $\mathcal{CU}[a,b]$ is a uniform distribution with the probability density function $f(x)=1/(b-a)\quad(a\leq x\leq b)$. During the model training, the loss function MSELoss is employed to quantify the discrepancy between predicted and real values, which has the characteristics of differentiability at any point and low computational complexity. According to the values of loss function, the gradients are calculated and backpropagated to an optimizer. An adaptive moment estimation (Adam) optimizer which combines the advantages of AdaGrad and RMSProp optimizers is chosen to update the weights \cite{adamoptimizer}. For this optimizer, $\eta$, $\beta_1$ and $\beta_2\in [0,1)$ are considered as the most important hyper-parameters. The $\beta_1$ and $\beta_2$ decide the exponential delay rates for the first moment and the second moment of the gradients, respectively, and the $\eta$ is usually set to a small constant, which can avoid zero denominator. The options of these hyper-parameters are $\beta_1=0.9$, $\beta_2=0.999$ and $\eta=10^{-8}$. The learning rate is also one of the most crucial hyper-parameters, which influences the convergence of the proposed model. Generally speaking, if a large learning rate is adopted, the model may not achieve better prediction accuracy, because it can make the weight keep on fluctuating around a reasonable value. Meanwhile, a small learning rate will lead to the difficulty of the model convergence. After many experiments, the learning rate is set to 0.0005.

\section{Performance Evaluation of the Proposed Model}
\subsection{Dataset Generation}
QuaDRiGa is a channel simulation platform recommended by the third generation partnership project (3GPP), which utilizes the geometry based stochastic modeling method to establish channel models, supporting three dimensional propagation, continuous time evolution and transitions between varying propagation scenarios \cite{quadriga1}. Considering a new mid band from 7 GHz and 24 GHz, also known as frequency range 3 (FR3), we simulate a time-varying wideband CF-mMIMO channel with the carrier frequency of 13 GHz. In this simulation, an 1.5 m-height UE moving in three scenarios including urban micro (UMi), urban macro (UMa) and rural macro (RMa) scenarios, and its moving speed varies in \{100, 150, 200, 250, 300\} km/h. Furthermore, this simulated channel consists of 64 APs with a single omnidirectional antenna distributed in the service areas with sizes of $250~m \times 250~m$, $500~m \times 500~m$ and $1000~m \times 1000~m$ randomly, and the height of APs is between 5 m and 25 m. To reduce the size of datasets and improve the training efficiency, the number of subcarriers is set to 16. Based on the channel simulation, we generate $10000 \times 16 \times 64$ CSI datasets of the time-varying wideband CF-mMIMO channel in each scenario, which are further divided into $8000 \times 16 \times 64$ training datasets and $2000 \times 16 \times 64$ testing datasets, respectively.

\subsection{Hyper-parameter Determination}

The hyper-parameters of the proposed model, such as the length of time window, the size of convolution kernel in the FreqConv layer, the adjacency matrix in the SpaceConv layer and so on, should be determined carefully, which will significantly influence the prediction accuracy.

\subsubsection{Length of Time Window} The length of time window refers to the number of consecutive time steps in the input sequence, which can be determined by time correlation analysis. The partial autocorrelation function (PACF) can be considered as an evaluation metric for time correlation analysis in this paper. Compared with autocorrelation function (ACF), the PACF can exclude the influence of indirect correlation and measure the direct correlation between current term and lag term. Moreover, it also avoids the multicollinearity to enhance the accuracy for selecting the length of time window. Supposed that a time sequence is $\begin{Bmatrix}x_1,x_2,\cdots,x_t\end{Bmatrix}$. The PACF between $x_t$ and $x_{t+k}$ can be calculated as
\begin{subequations}
  \begin{align}
      &\phi_k=\frac{\gamma_k-\sum_{i=1}^{k-1}\phi_i\gamma_{k-i}}{\gamma_0-\sum_{i=1}^{k-1}\phi_i\gamma_{k-i}}\quad (k\geq2)\text{,}\\
      &s.t.\quad \phi_1=\gamma_1/\gamma_0 \text{,}
  \end{align}
\end{subequations}
where $\gamma_i$ is the autocovariance between $x_t$ and $x_{t+i}$, and $k\in \mathbb{N}^*$ indicates the lag value. 

\begin{figure}[!t]
\centering
\includegraphics[trim=5pt 0pt 5pt 15pt,clip,scale=0.56]{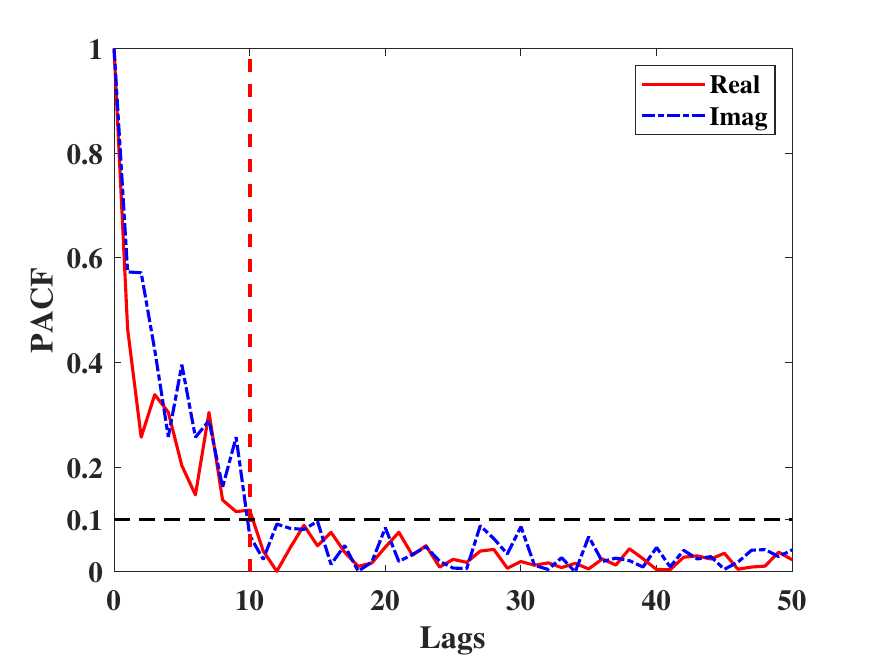}
\caption{The PACF results of CSI. \label{fig:3}} 
\end{figure}

\begin{figure}[!t]
\centering
\includegraphics[trim=5pt 0pt 5pt 15pt,clip,scale=0.56]{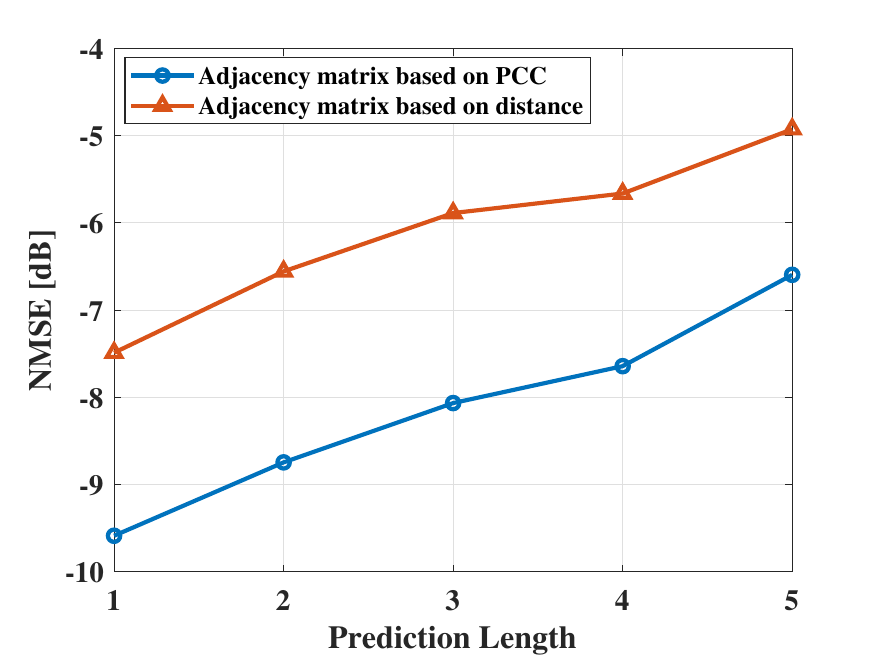}
\caption{Comparison of NMSE using different adjacency matrices. \label{fig:4}} 
\vspace{-0.45cm}
\end{figure}

Fig.~\ref{fig:3} illustrates the PACF results of the real and imaginary parts of datasets. It can be seen that with the gradual increase of the lag value, the PACF values decrease rapidly. When the lag value is 10, the PACF values drop below 0.1, which demonstrates that there is almost no correlation between the current term and the lag term at this point. Hence, the length of time window should be set to 10.

\subsubsection{Adjacency Matrix} The adjacency matrix encodes the topological relationships between nodes, typically constructing by edge weight. In this paper, we consider two different methodologies for adjacency matrix construction. The one is the distance-based construction method. Specifically, we compute the distances between different nodes, and derive the matrix weights using an exponential decay funtion to generate the elements of the adjacency matrix, expressed as
\begin{equation}
  \label{eq:adjacency1}
  a_{ij}=e^{-\frac{\left|s_{ij}\right|}{\sigma}} \text{,}
\end{equation}
where $s_{ij}$ represents the distance between AP $i$ and AP $j$, $i,j\in \{1,2,\cdots,M\}$, $a_{ij} \in [0,1]$, and $\sigma$ is an adjustable parameter which is used to control the attenuation rate of $a_{ij}$ with the distance $s_{ij}$ increasing. According to Eq.~\eqref{eq:adjacency1}, we find that the elements $a_{ij}$ decrease monotonically as the distance between the APs increases. However, it only focuses on the distance between APs, and does not pay attention to the linear correlation between different APs. 

\begin{figure}[!t]
\centering
\includegraphics[trim=30pt 0pt 5pt 15pt,clip, scale=0.6]{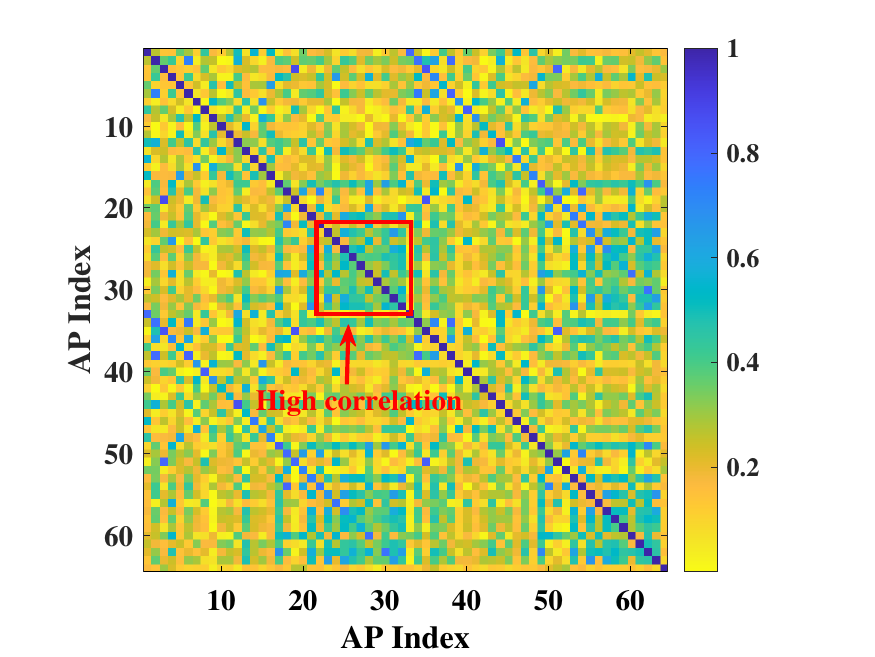}
\caption{Adjacency Matrix. \label{fig:PCC邻接矩阵图}} 
\vspace{-0.3cm}
\end{figure}

\begin{figure}[!t]
\centering
\includegraphics[trim=0pt 0pt 15pt 15pt,clip, scale=0.61]{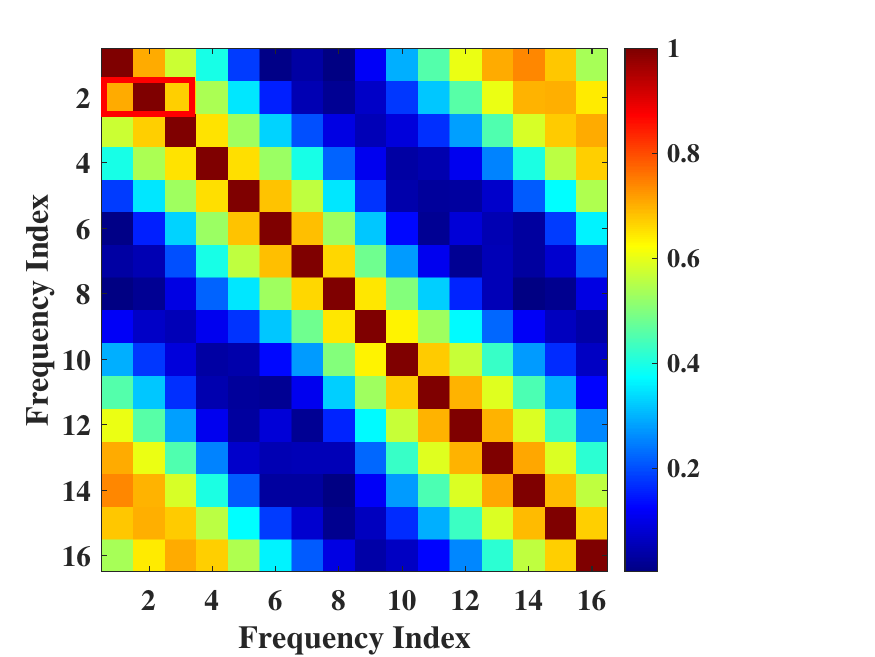}
\caption{The results of PCC between subcarriers. \label{fig:频域PCC图}} 
\vspace{-0.45cm}
\end{figure}

The other is the Pearson correlation coefficient (PCC)-based construction method, which considers the linear correlation between different APs. The PCC serves as a quantitative measurement of linear correlation between different variables, which is between $-$1 and 1. The larger the absolute value of PCC is, the higher the correlation is. In this method, we calculate the PCC between APs, and use the absolute value to obtain the elements of this adjacency matrix, shown as
\begin{equation}
  a_{ij} = |r_{ij}| \text{,}
\end{equation}
where $r_{ij}$ represents the PCC between AP $i$ and AP $j$

We compare the NMSE results of the proposed model with the two adjacency matrices, as illustrated in Fig.~\ref{fig:4}. The experimental results demonstrate the superior prediction accuracy when using the PCC-based construction method. This is because the adjacency matrix generated by the distance-based construction method cannot accurately reflect the correlation between APs. For example, when one AP is near from the other AP, but the PCC value between them is too small. In this case, irrelevant correlation information contained in the distance-based adjacency matrix is input into the proposed model, thereby deteriorating the prediction accuracy. Thus, the PCC-based construction method is recommended in this paper, and the adjacency matrix generated by this method is shown in Fig.~\ref{fig:PCC邻接矩阵图}. It can be seen that the correlation results between different APs are greater than 0.6 in a red box, which can be regarded as a relatively high correlation between some APs.

\begin{figure}[!t]
\centering
\includegraphics[trim=5pt 0pt 5pt 15pt,clip,scale=0.56]{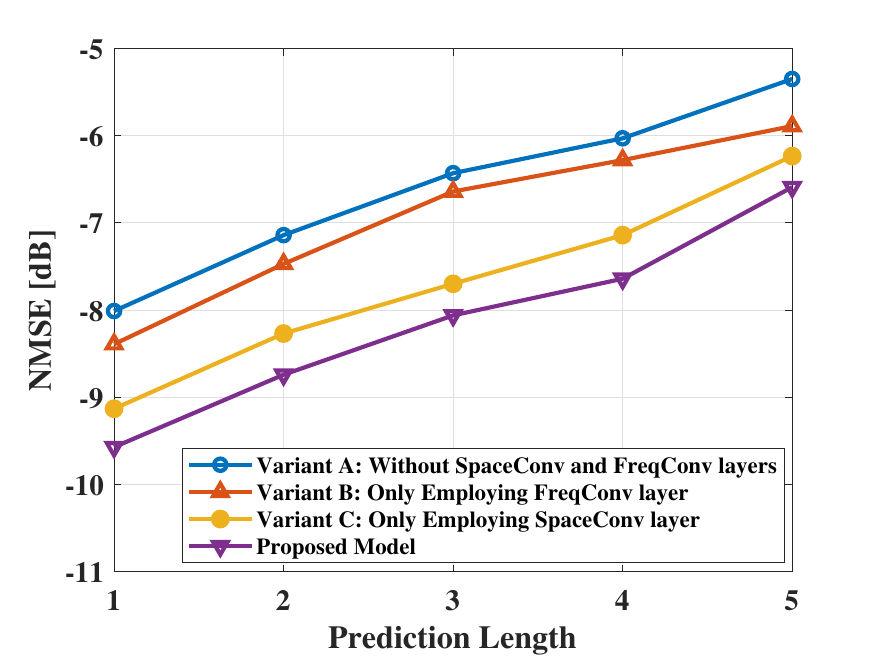}
\caption{Ablation Experiment. \label{fig:6}}
\vspace{-0.45cm}
\end{figure}  

\subsubsection{Size of Convolution Kernel}The size of convolution kernel is one of the most important hyper-parameters in the FreqConv layer, which needs to be meticulously chosen by the PCC between subcarriers. If the convolution kernel is too small, it may fail to capture sufficient information in the frequency domain. If the convolution kernel is too large, it might extract the low correlation in the frequency domain, thereby deteriorating the prediction accuracy. As shown in Fig.~\ref{fig:频域PCC图}, it is observed that there is high correlation between each subcarrier and its adjacent subcarriers. In other words, the local correlation is high but the global correlation is low in the frequency domain. The convolution kernel in the FreqConv layer should cover a certain region with the high frequency correlation. Therefore, according to the results of frequency correlation analysis, the size of convolution kernel should be set to $1 \times 3$, which is able to cover the region with the highest correlation in the frequency domain.

\subsubsection{Other Hyper-Parameters}We use 5-fold cross-validation method to choose the other hyper-parameters, including the size of feature dimensions, the number of neurons in the dense sub-layer and the dimension of query and key matrices. The 5-fold cross-validation can be realized in three steps: firstly, the datasets are divided into five parts equally, where four parts are used as the training datasets and one part is used as the testing datasets. Then, we train the proposed model and obtain the prediction accuracy. Finally, the above steps are repeated five times and the average accuracy is calculated, which is served as the basis for hyper-parameter selection. Following this method, the size of input dimensions and the number of neurons in the Transformer-encoder layer are configured to 128, and the dimension of query and key matrices is set to 64.

\subsection{Prediction Accuracy} 
In this subsection, the effectiveness of the SpaceConv and FreqConv layers is verified by ablation experiment. Subsequently, the prediction accuracy of the proposed model is compared with that of traditional DL models. Finally, the influences of space-time-frequency correlations on prediction accuracy are investigated. 

\begin{figure}[!t]
\centering
\includegraphics[trim=5pt 0pt 5pt 15pt,clip,scale=0.56]{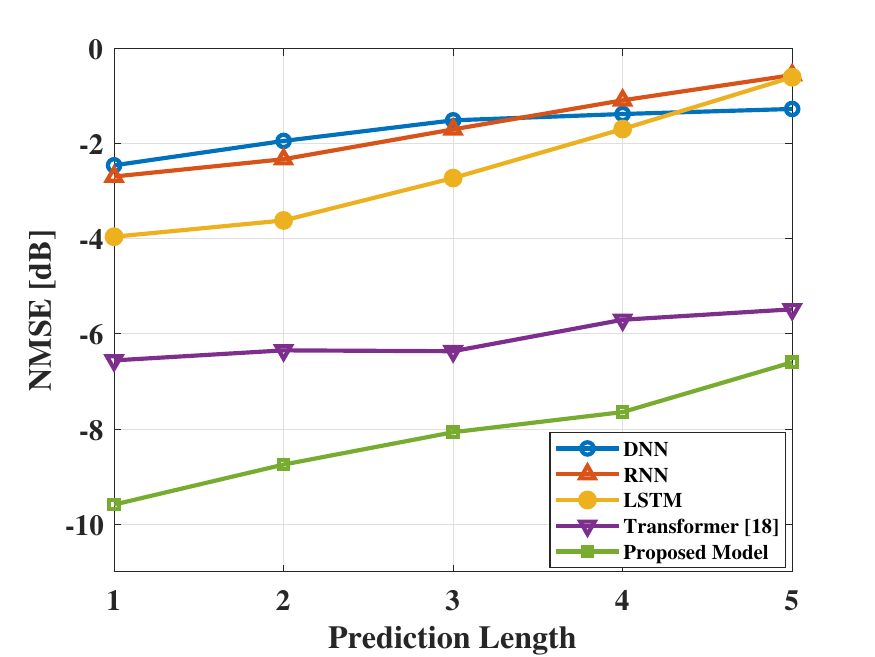}
\caption{Comparison of the prediction accuracy between the proposed model and traditional DL models. \label{fig:7}}
\vspace{-0.45cm}
\end{figure} 

\subsubsection{Ablation Experiment} Ablation experiments are conducted by gradually removing several components to assess their impact on prediction accuracy and verify their effectiveness. In this paper, the SpaceConv and FreqConv layers are gradually removed to obtain three variants of the proposed model: without SpaceConv and FreqConv layers (Variant A), only employing SpaceConv layer (Variant B), and only employing FreqConv layer (Variant C). Fig.~\ref{fig:6} compares the prediction accuracy of the proposed model and its three variants in UMi LoS scenario, where the UE moves with the speed of 100 km/h in the $250~m \times 250~m$ service area. It can be seen that both Variant B and Vairant C achieve higher prediction accuracy compared with Variant A. Notably, the proposed model further surpasses both Variant B and Variant C, achieving the highest prediction accuracy among all variants. These results mean that the SpaceConv and FreqConv layers are able to improve the prediction accuracy independently without mutual interference. Furthermore, the results show that the NMSE of Variant B is approximately 0.4 dB higher than that of Variant A, while the NMSE of Variant C exhibits a larger improvement, being roughly 1 dB higher than Variant A. This indicates that the SpaceConv layer contributes to a more significant performance improvement in the model compared to the FreqConv layer.

\subsubsection{Results of Comparison with Different DL Models} We utilize the same datasets as in ablation experiment to compare the prediction accuracy of the proposed model and traditional DL models, such as deep neural network (DNN), RNN, LSTM and Transformer \cite{10480335}, as shown in Fig.~\ref{fig:7}. This improvement in prediction accuracy stems from the ability of the proposed model for capturing the space-time-frequency correlations, whereas traditional deep learning models rely solely on time correlation, limiting their prediction accuracy. The Transformer model demonstrates superior prediction accuracy compared with both RNN and LSTM networks. Since the proposed model and Transformer outputs multi-step prediction results in parallel through a dense layer, it avoids the degradation of prediction accuracy caused by error propagation between cells in the RNN and LSTM networks. Furthermore, the DNN demonstrates the lowest prediction accuracy among the evaluated models. This is primarily due to its simplistic architecture, it inherently limits its capacity to capture time correlations effectively. The prediction accuracy of RNN and LSTM is higher than that of DNN, especially when the prediction length is relatively small. For all models, the prediction accuracy decreases as the prediction length increases. The key factor lies in the stochastic sampling of channel parameters, inducing heightened randomness in this generated channel and diminishing the predictability.

\begin{table*}[!t]
\renewcommand\arraystretch{1.4}
\small \caption{Comparison of computational complexity of multiple models} \label{table2} \centering
\begin{tabular}{|c|c|c|c|c|}
\hline
\multirow{2}{*}{\textbf{Model}} & \multicolumn{2}{c|}{\textbf{Space complexity}} & \multicolumn{2}{c|}{\textbf{Time complexity}}\\
\cline{2-5}
& \textbf{Number of parameters} & \textbf{Memory usage (MB)} & \textbf{Number of FLOPs} & \textbf{Computational time (s)}\\
\hline
DNN & $1.25\times 10^6$ & 4.77 & $2.49\times 10^6$ & $1.59 \times 10^{-7}$\\
\hline
RNN & $1.28\times 10^6$ & 4.88 & $2.56\times 10^6$ & $1.63 \times 10^{-7}$\\
\hline
LSTM & $4.37\times 10^6$ & 16.67 & $8.75\times 10^6$ & $5.57 \times 10^{-7}$\\
\hline
Transformer \cite{10480335} & $2.87\times 10^6$ & 10.95 & $4.26\times 10^7$ & $2.71 \times 10^{-6}$\\
\hline
Proposed Model & $1.46\times 10^6$ & 5.57 & $2.83\times 10^7$ & $1.80 \times 10^{-6}$\\
\hline
\end{tabular} 
\end{table*}

\begin{figure}[!t]
\centering
\includegraphics[trim=5pt 0pt 5pt 15pt,clip,scale=0.56]{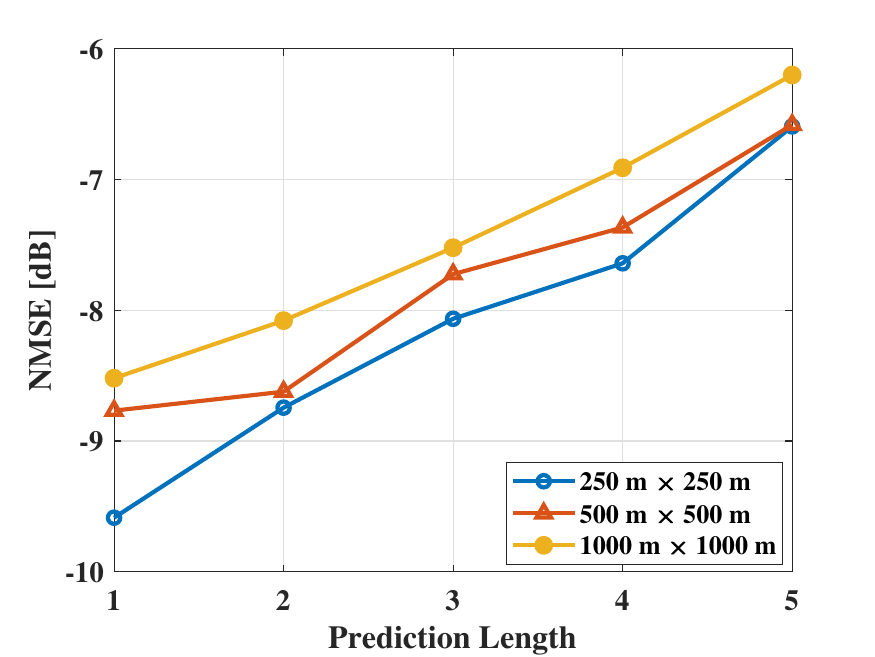}
\caption{Comparison of prediction accuracy in the service areas with different sizes. \label{fig:不同空间}}
\end{figure} 

\subsubsection{Impact of Space-Time-Frequency Correlations on Results} To thoroughly investigate the impacts of space-time-frequency correlations on prediction accuracy, we conduct the extensive evaluations across diverse conditions, such as different sizes of service areas, varying UE velocities and distinct scenarios. 
\begin{figure}[!t]
  \centering
  \includegraphics[trim=5pt 0pt 5pt 15pt,clip,scale=0.56]{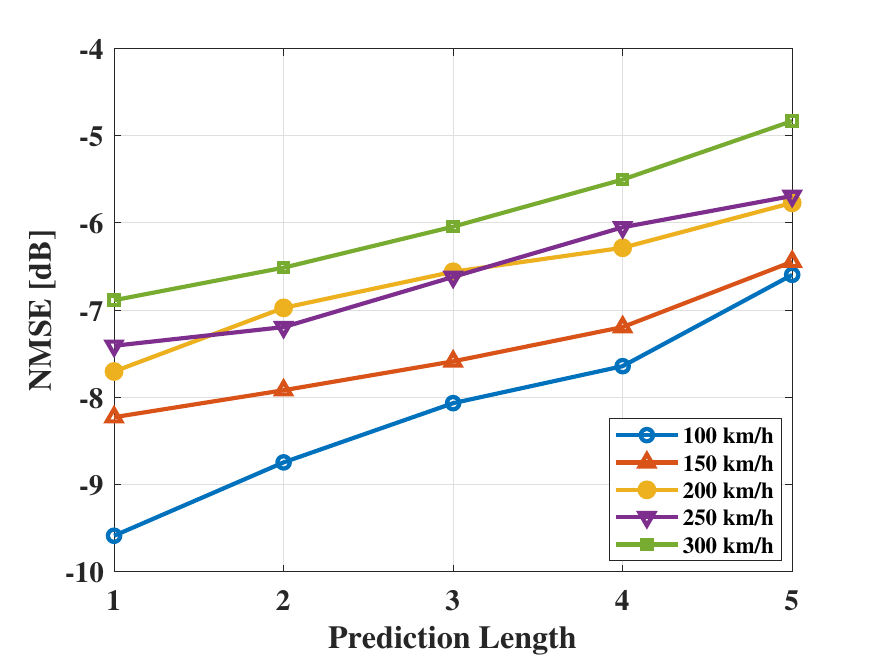}
  \caption{Comparison of prediction accuracy in different UE velocities. \label{fig:9}}
  \vspace{-0.45cm}
  \end{figure} 
  
  \begin{figure}[!t]
  \centering
  \includegraphics[trim=5pt 0pt 5pt 15pt,clip,scale=0.56]{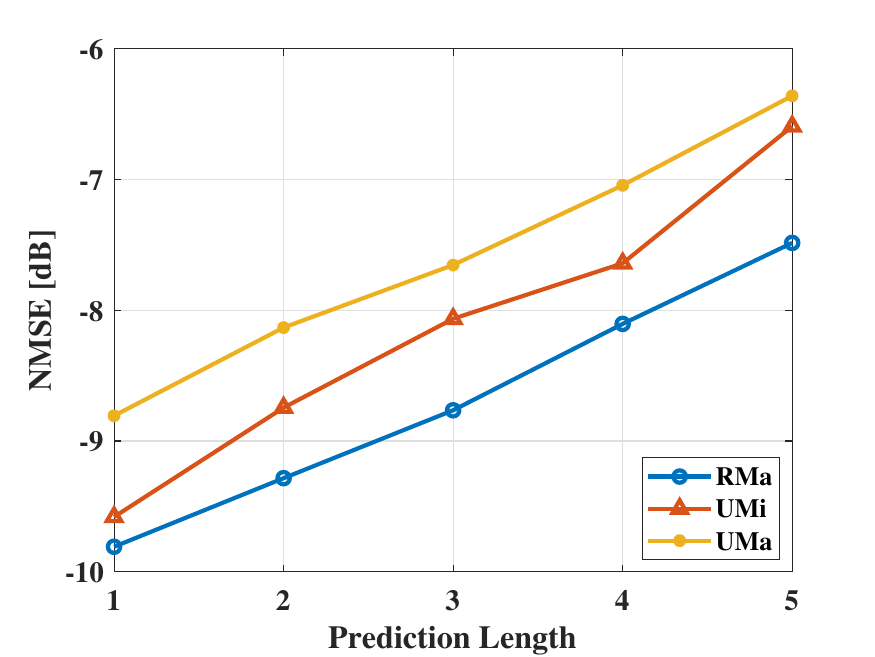}
  \caption{Comparison of prediction accuracy in different scenarios. \label{fig:10}}
  \vspace{-0.45cm}
  \end{figure}

The size of service areas is one of factors influencing the space correlation. Specifically, the expansion of service areas leads to an increase in the average distance between APs, which consequently attenuates the space correlation. Fig.~\ref{fig:不同空间} shows the NMSE results in the service areas with different sizes in the UMi scenario when the UE velocity is 100 km/h. The experimental results demonstrate a consistent degradation in the prediction accuracy of the proposed model as space correlation diminishes. When the space correlation is low, the SpaceConv layer cannot extract sufficient effective information, thus deteriorating the prediction performance. Further analysis, the prediction accuracy in the $1000~m \times 1000~m$ service area decreases by approximately 1 dB compared to that in the $250~m \times 250~m$ service area, which is roughly equal to the difference in prediction accuracy between Variant A and Variant C in the ablation experiment. When APs are distributed within the $1000~m \times 1000~m$ service area, there is almost no correlation between them. This indicates that the space correlation boosts NMSE of the proposed model by approximately 1 dB.

As the UE velocity increases, the Doppler shift grows larger and the Doppler spread widens, leading to a corresponding reduction in time correlation. Fig.~\ref{fig:9} compares the NMSE results when the UE moves at different velocities in UMi scenarios. It is observed that the proposed model exhibits a decline in prediction accuracy as the time correlation decreases. When the length of time window remains unchanged, the Transformer-encoder layer struggles to establish meaningful long-range correlations, which leads to reduce the accuracy in sequence prediciton. Therefore, in order to maintain the high prediction accuracy, the length of time window should be appropriately adjusted when the UE moves at ultra-high speeds, such as 1000 km/h, 2000 km/h and so on.

\begin{figure*}[!t]
\centering
\includegraphics[trim=15pt 15pt 15pt 15pt,clip, scale=0.39]{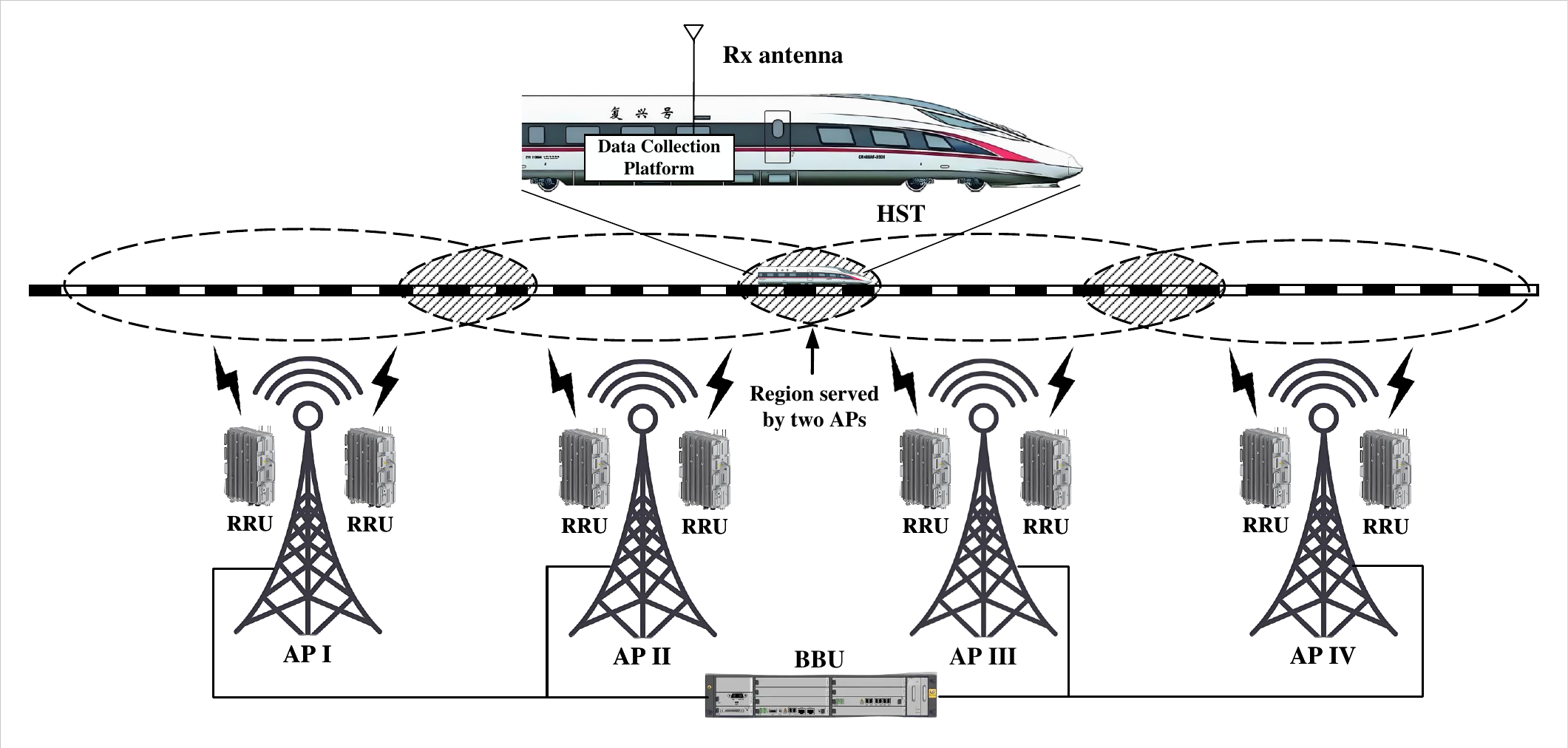}
\caption{The structure of the measurement system. \label{fig:11}} 
\vspace{-0.3cm}
\end{figure*} 

The strength of frequency correlation is determined by the magnitude of the delay spread (DS). A larger delay spread leads to more pronounced frequency-selective fading, thereby decreasing the frequency correlation. When we assess the impact of frequency correlation on prediction accuracy, we compares the NMSE of the proposed model in different scenarios with different DS in the $250~m \times 250~m$ service area, such as UMi, UMa and RMa scenarios, as shown in Fig.~\ref{fig:10}. Note that the velocity of UE is set to 100 km/h. It is observed that the prediction accuracy of the proposed model decreases slightly with the DS increasing. Under this situation, convolution kernel in the FreqConv layer will extract the redundant information, since it does not cover the region with the highest correlation in the frequency domain. Hence, when the frequency correlation is higher or lower, we should adjust the size of the convolution kernel to maintain the high prediction accuracy.

\subsection{Computational Complexity}
The computational complexity includes both space and time complexity, which can be shown as the number of trainable parameters and floating-point operations per second (FLOPs). The number of trainable parameters refers to the number of weights, which are continuously adjusted during the model training. Likewise, the number of FLOPs means that the total number of basic arithmetric operations performed within the model, which includes addition, subtraction, multiplication, division, exponentiation and square root. Based on the number of parameters and FLOPs, the size of memory usage and computational time can be further obtained, as follows
\begin{equation}
  Memory=\frac{4N_{\text{parameters}}}{1024^2}~~(\text{MB})\text{,}
\end{equation}
\begin{equation}
  Time=\frac{FLOPs}{f_{\text{GPU}}N_{\text{Unit}}N_{\text{Core}}}\text{,}
\end{equation}
where $N_{\text{parameters}}$ represents the number of trainable parameters, $f_{\text{GPU}}$ is the core frequency of the graphics processing unit (GPU), $N_{\text{Unit}}$ stands for the number of stream processors, and $N_{\text{Core}}$ denotes the number of compute unified device architecture (CUDA) cores.

Using these two metrics, the computational complexity of the proposed model is analyzed and compared with traditional DL models. When we analyze the computational complexity of DL models, the layer number of DNN, RNN and LSTM is set to two, and the settings of hyper-parameters in Transformer are the same as those in the Transformer-encoder layer of the proposed model. Table~\ref{table2} lists the results of computational complexity for the DNN, RNN, LSTM, Transformer \cite{10480335} and the proposed model. It can be seen that the number of trainable parameters and FLOPs are reduced compared with the Transformer, since the proposed model uses only the Transformer-encoder and replacing traditional convolution with DSC. Moreover, the number of parameters and FLOPs of the LSTM are approximately four times as large as those of RNN. This is because the LSTM contains three gated units and one memory cell, which leads to a significant increase in memory usage and computational time. Additionally, the number of parameters in the Transformer and the proposed model is reduced but the number of FLOPs is increased compared with that of the LSTM.

In this study, these DL models are trained on an NVIDIA V100 GPU equipped with 5120 CUDA cores and a base clock frequency of 1350 MHz. Experimental results show that the proposed model achieves a computational latency of $1.80\times 10^{-6}$ seconds for CSI prediciton when the prediction length is five. Such efficiency underscores the capability of the proposed model for real-time inference, making it suitable for rapid time-varying channels.
  
\section{Validation of the Proposed Model in High-Speed Railway Scenarios}

\subsection{Data Collection}
In order to verify the prediction accuracy of the proposed model, we collected the realistic channel data on Beijing to Tianjin (BT) high-speed railway (HSR) in China when the HST operated at a velocity of 285 km/h. \cite{7277055}. The used measurement system consists of a data collection platform and a HST LTE network with the architecture of building baseband unit (BBU) plus remote radio unit (RRU), as shown in Fig.~\ref{fig:11}. The data collection platform connects a omnidirectional train-mounted antenna, which enables to collect the LTE cell-specific reference signal (CRS) with the carrier frequency of 1.89 GHz and the bandwidth of 18 MHz. In this HST LTE network, several APs are approximately 30 m away from the rail with the spacing of 1.2 km and the height of 30 m. Additionally, two RRUs are deployed in each AP, which transmit signals by directional antennas in opposite directions along the rail. In this environment, the transmit antennas are deployed at a height significantly exceeding the surrounding railway environment, which primarily consisted of sparse vegetation and low-rise structures with an average elevation below 10 m. During the data collection, we mainly focus on the channel data of regions served by each two APs and obtain the CSI in the next subsection.

\begin{figure}[!t]
\centering
\includegraphics[trim=15pt 15pt 15pt 15pt,clip, scale=0.6]{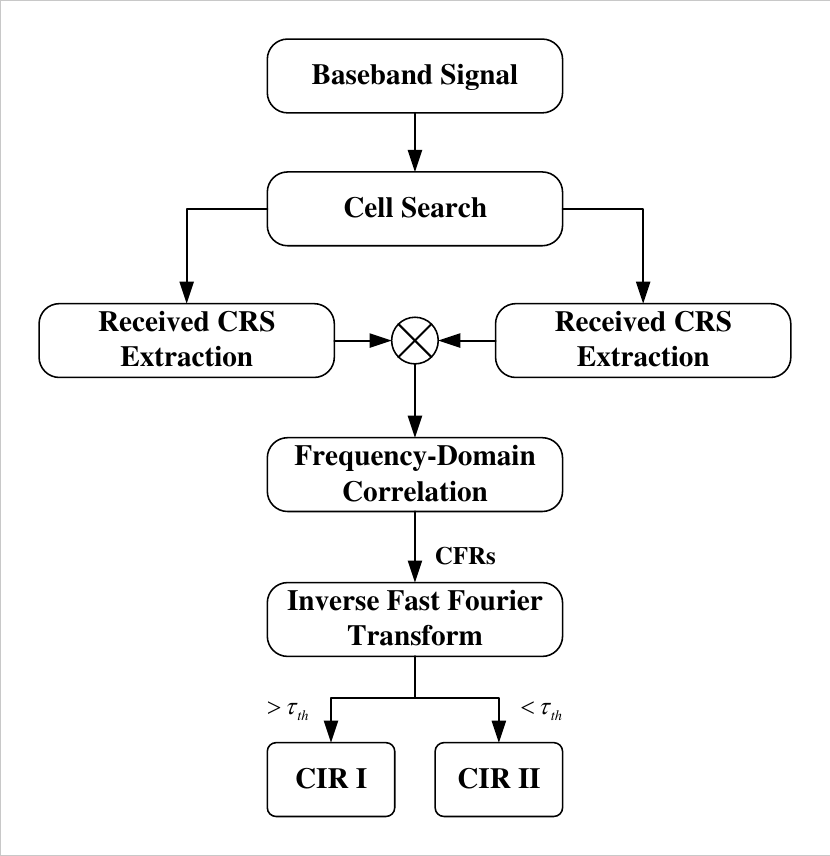}
\caption{The process flow for realistic data. \label{fig:12}} 
\vspace{-0.45cm}
\end{figure}

\subsection{Data Processing and Analysis}
\subsubsection{Data Processing} The baseband signal data obtained in the regions served by each two APs is processed offline. The data process flow is illustrated in Fig.~\ref{fig:12}, and the main steps are described as follows
\begin{itemize}
  \item The cell identification (ID) is determined through the cell research, and synchronized frames are acquired to perform the local CRSs generation and received CRSs extraction. 
  \item The channel frequency responses (CFRs) are derived by the frequency-domain correlation of generated and received CRSs locally.
  \item The CFRs are converted to channel impulse responses (CIRs) through inverse fast Fourier transform (IFFT), where the power delay profile (PDP) is subsequently calculated. 
  \item A time-delay-window-based partitioning method \cite{8493616} with a reasonable threshold $\tau_{th}$ is employed to isolate the CIRs from different APs. 
\end{itemize}

\begin{figure}[!t]
\centering
\includegraphics[trim=5pt 0pt 5pt 15pt,clip, scale=0.56]{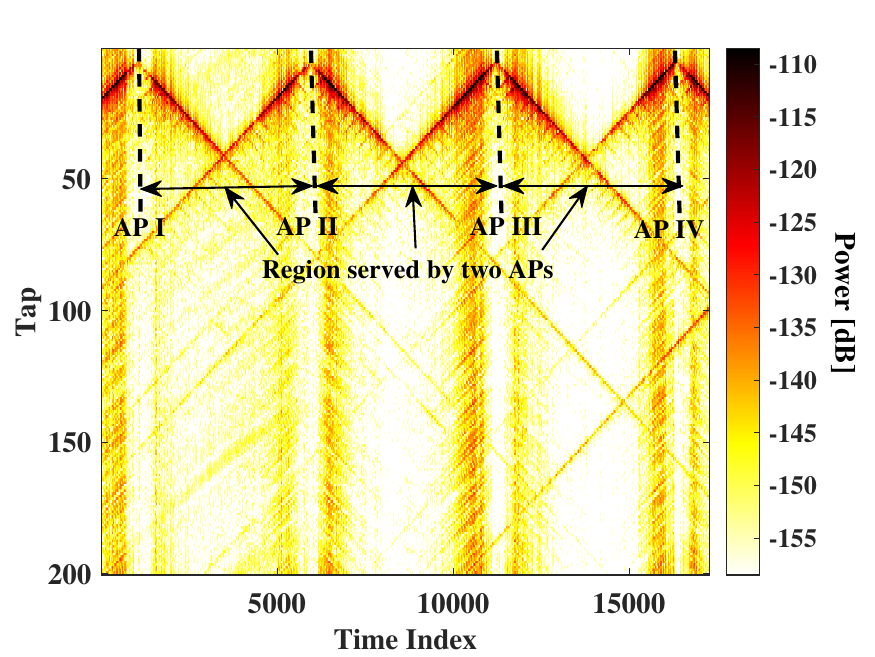}
\caption{PDP of realistic data. \label{fig:PDP}} 
\end{figure}

\begin{figure}[!t]
\centering
\includegraphics[trim=5pt 0pt 5pt 15pt,clip, scale=0.56]{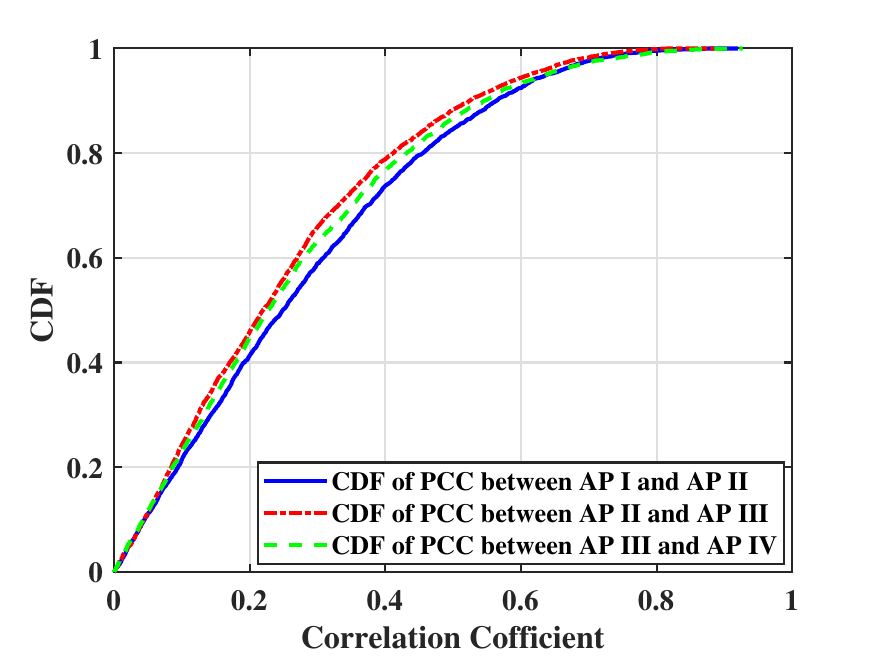}
\caption{The CDF of PCC between different APs. \label{fig:14}} 
\vspace{-0.45cm}
\end{figure}

As illustrated in Fig.~\ref{fig:PDP}, the proposed data processing pipeline generates the PDP results from the collected channel data. The analysis reveals three regions served by each two APs, where the maximum time delay of most effective multipath components (MPCs) is less than $1~\mu s$ in this scenario. Based on this observation, we empirically set the threshold $\tau_{th}=1$ to obtain the CIRs of two APs. After that, the fast Fourier transform (FFT) is applied to convert the CIRs into $9252 \times 18 \times 2$ CFR datasets. These datasets are further partitioned into $7401 \times 18 \times 2$ training datasets and $1851 \times 18 \times 2$ testing datasets.

\begin{figure}[!t]
\centering
\subfloat[]{
\includegraphics[trim=15pt 0pt 700pt 0pt,clip,scale=0.2]{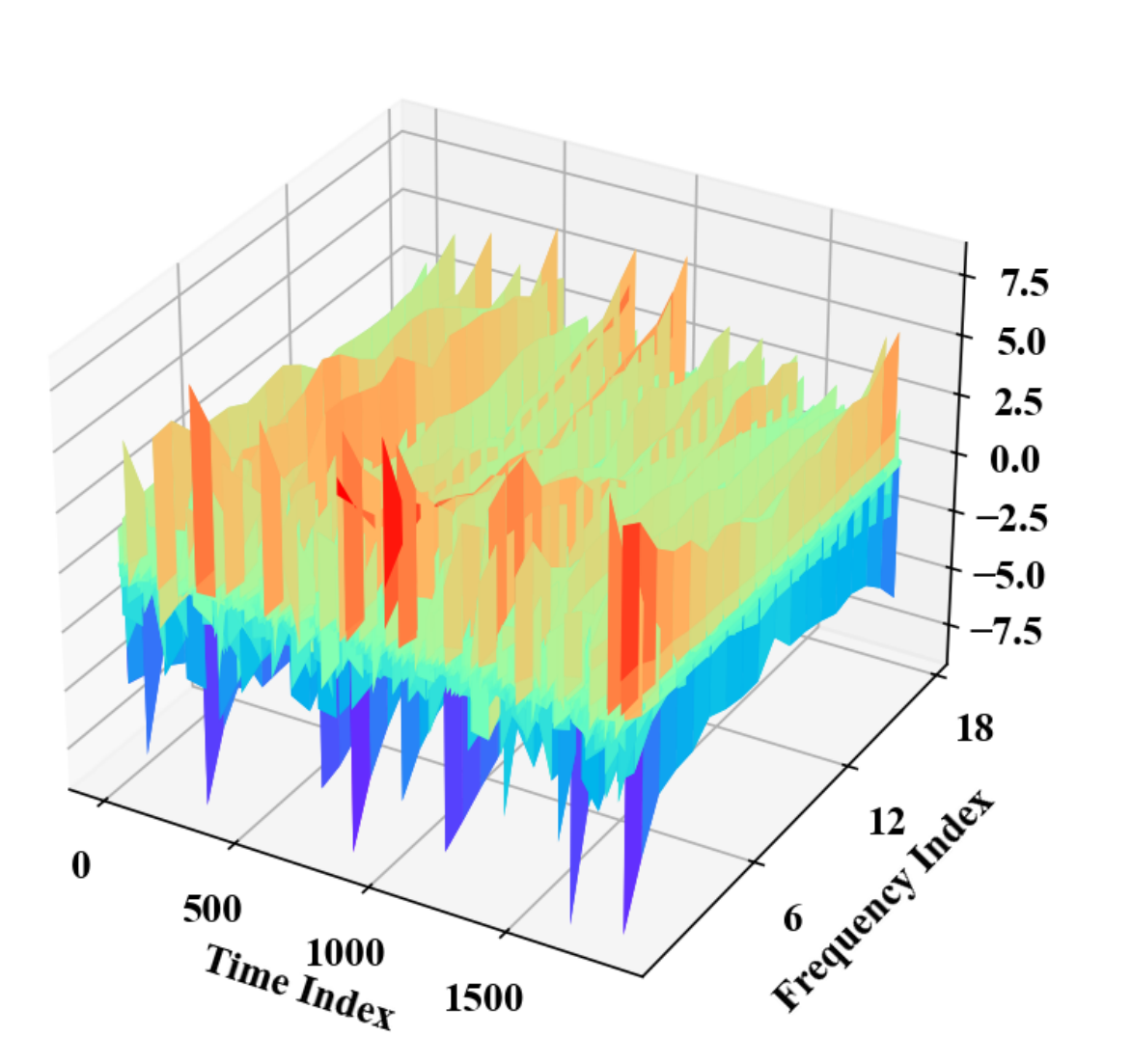}
}
\subfloat[]{
\includegraphics[trim=650pt 0pt 10pt 0pt,clip,scale=0.2]{Comparison_real_predicted1-eps-converted-to.pdf}
}
\\
\subfloat[]{
\includegraphics[trim=15pt 0pt 700pt 0pt,clip,scale=0.2]{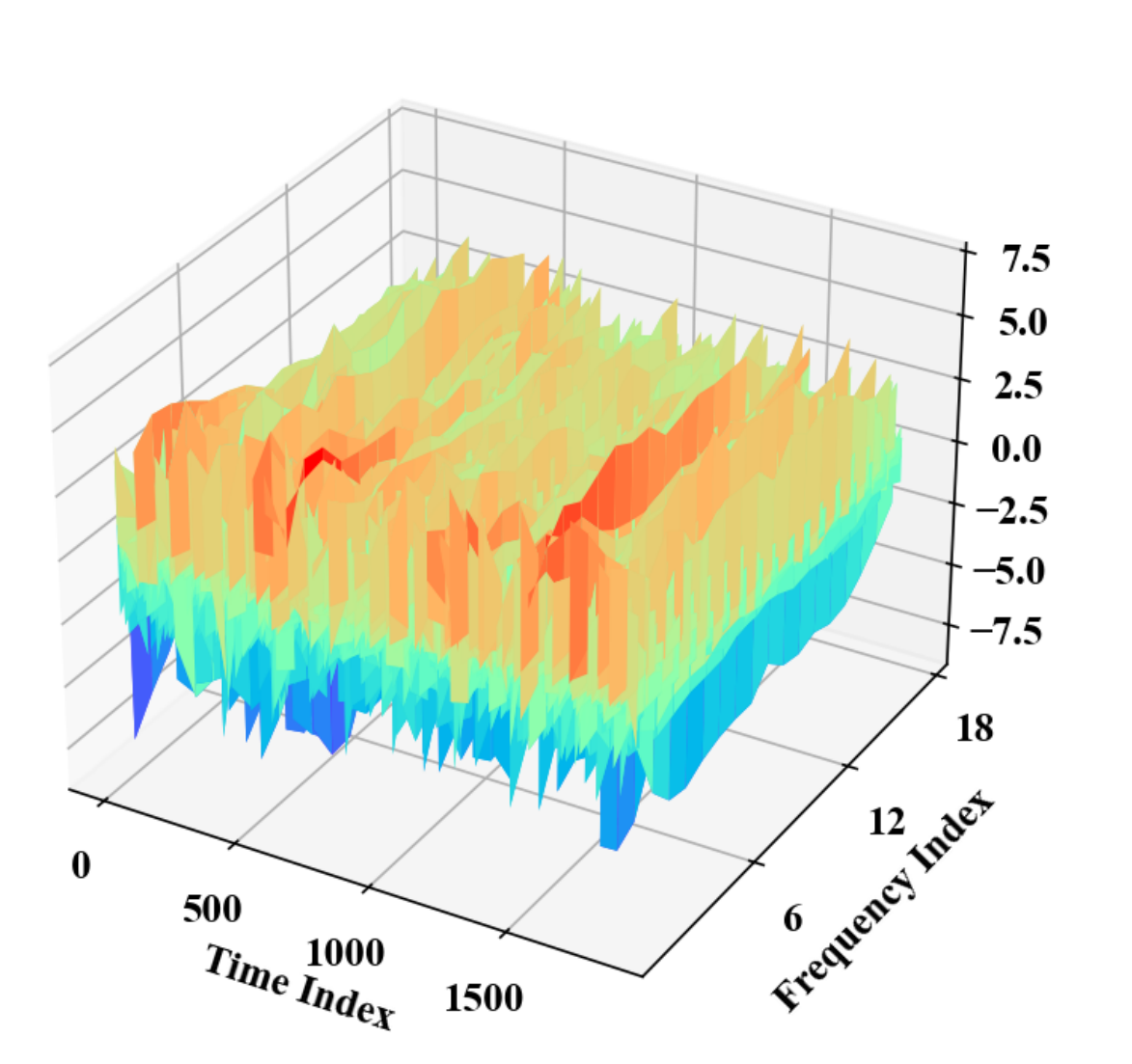}
}
\subfloat[]{
\includegraphics[trim=650pt 0pt 10pt 0pt,clip,scale=0.2]{Comparison_real_predicted2-eps-converted-to.pdf}
}
\caption{Comparison of predicted and actual results of CSI for two APs. (a) Predicted value for CSI of one AP. (b) Actual value for CSI of one AP. (c) Predicted value for CSI of other AP. (d) Actual value for CSI of other AP. \label{fig:实测真实值预测值对比1}}
\vspace{-0.45cm}
\end{figure} 

\subsubsection{Analysis for Space-Time-Frequency Correlations} As the HST moves, the rapid changes of scenarios induce significant variations in space-time-frequency correlations. For the space correlation analysis, we compute the PCC $c$ between CSI at each time index, and utilize its cumulative distribution function (CDF) $\mathrm{F}_\mathrm{c}(c)$ to analyze the space correlation, illustrated in Fig.~\ref{fig:14}. It can observed that more than approximate 80\% of the PCC values are less than 0.4. This statistically confirms low space correlation between APs, attributable to their large inter-AP distance. Based on these results, the mean PCC value of 0.3 is adopted to construct the adjacency matrix. After that, for time correlation analysis, we calculate the PACF initialized at each time index, obtain its mean value to yield an optimal time window length of five. Similarly, the PCC at each time index is calculated in the frequency domain, and the size of convolution kernel is set to $1 \times 3$ according to its mean value.

\subsection{Validation Results}

The prediction accuracy of the proposed model is validated using realistic datasets measured in HST LTE networks. Fig.~\ref{fig:实测真实值预测值对比1} shows the comparison of the predicted and actual values of the CSI based on realistic datasets. It can be seen that the predicted value can fit the actual value well, but minor deviations occur at peak points characterized by rapid fluctuation. On the whole, it shows that the proposed model is capable of capturing the space-time-frequency correlations to grasp the changing trend of CSI, thus leading to the high prediction accuracy. 

\begin{figure}[!t]
\centering
\includegraphics[trim=5pt 0pt 5pt 15pt,clip,scale=0.56]{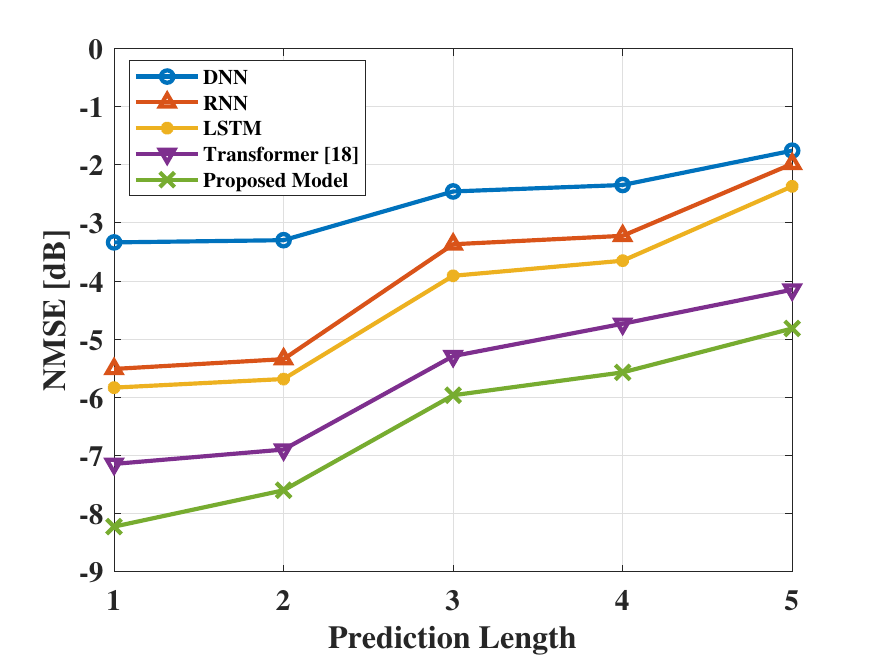}
\caption{Comparison of prediction accuracy for realistic datasets. \label{fig:15}}
\vspace{-0.45cm}
\end{figure}

The NMSE are used to compare the prediction accuracy of the proposed model and traditional DL models based on realistic datasets, which are shown in Fig.~\ref{fig:15}. It can be seen that the proposed model is superior to other traditional models by NMSE in HST LTE networks. We further discovered that the proposed model achieves an average improvement of approximately 2 dB in NMSE compared with the Transformer across all prediction steps. In contrast, the proposed model achieves only an approximately 1 dB improvement in NMSE in HST LTE networks. This is because the spacing of APs is 1.3 km in the HST LTE network, which is significantly larger than that in the simulation scenario. This is becasuse the enlarged AP spacing leads to low space correlation, consequently diminishing the extent of the improvement in prediction accuracy. However, as shown in Table~\ref{table2}, the proposed model exhibits the lower computational complexity compared with the traditional Transformer. This advantage renders it particularly suitable for real-time applications in HST LTE networks.

\section{Conclusion}
This paper presents a investigation of DL-based channel prediction for CF-mMIMO systems, explicitly incorporating joint space-time-frequency correlations to achieve accurate CSI acquisition. First of all, the prediction problems have been formulated, which can predict the multi-step CSI without error propagation. Then, a novel model has been proposed for CF-mMIMO systems with irregular AP deployment, which utilizes the space-time-frequency correlations and extracts the space correlation in irregular AP deployment. Next, QuaDRiGa platform has been used to generate the CSI datasets under different sizes of service areas, UE velocities and scenarios. Correlation analysis and cross-validation have been implemented to determine the optimal hyper-parameters. The performance of the proposed model has been evaluated by the prediction accuracy and computational complexity. It has been shown that the proposed model has the high prediction accuracy, and it has lower computational complexity compared with the vanilla Transformer. After that, the impacts of space-time-frequency correlations on prediction accuracy are studied. Finally, the realistic datasets obtained by the HST LTE networks have been used to validate the performance. It has been shown that the proposed model has the best prediction performance in HST LTE networks compared with traditional DL models.

\bibliographystyle{IEEEtran}
\bibliography{ref.bib}

\end{document}